\documentclass[aps,prd,twocolumn,amsmath,showkeys, superscriptaddress]{revtex4}

\usepackage{graphicx}
\usepackage{longtable}
\usepackage{float}
\usepackage{dcolumn}
\usepackage{bm}
\usepackage{amsfonts}
\usepackage{appendix}
\usepackage{multirow}
\usepackage{color}
\usepackage{soul}
\usepackage[normalem]{ulem}
\usepackage{amsmath}
\usepackage{cancel}
\usepackage{url}
\usepackage{footnote}

\newcommand{\be}{\begin{equation}}
\newcommand{\ee}{\end{equation}}
\newcommand{\bea}{\begin{eqnarray}}
\newcommand{\eea}{\end{eqnarray}}
\def\ba{\begin{array}}
\def\ea{\end{array}}

\begin{document}

\preprint{APS/123-QED}

\title{Comparative analysis of local angular rotation between the Ring Laser Gyroscope GINGERINO and GNSS stations}

\author{Giuseppe Di Somma}
\email{g.disomma1@studenti.unipi.it}
\affiliation{ Dipartimento di Fisica,   Universit\`a di Pisa, largo B. Pontecorvo 3, I-56127 Pisa, Italy,}
\affiliation{Istituto Nazionale di Fisica Nucleare (INFN), sez. di Pisa, largo B. Pontecorvo 3, I-56127 Pisa, Italy.}

\author{Nicolò Beverini}
\affiliation{ Dipartimento di Fisica,   Universit\`a di Pisa, largo B. Pontecorvo 3, I-56127 Pisa, Italy,}

\author{Giorgio Carelli}
\affiliation{ Dipartimento di Fisica,   Universit\`a di Pisa, largo B. Pontecorvo 3, I-56127 Pisa, Italy,}

\author{Simone Castellano}
\affiliation{Istituto Nazionale di Fisica Nucleare (INFN), sez. di Pisa, largo B. Pontecorvo 3, I-56127 Pisa, Italy.}
\affiliation{Gran Sasso Science Institute,  Viale Francesco Crispi 7, 67100 L'Aquila AQ, Italy}

\author{Roberto Devoti}
\affiliation{Istituto Nazionale di Geofisica e Vulcanologia, Sez. ONT, sede di Roma, Italy}
 
\author{Enrico Maccioni}
\affiliation{ Dipartimento di Fisica,   Universit\`a di Pisa, largo B. Pontecorvo 3, I-56127 Pisa, Italy,}
\affiliation{Istituto Nazionale di Fisica Nucleare (INFN), sez. di Pisa, largo B. Pontecorvo 3, I-56127 Pisa, Italy.}

\author{Paolo Marsili}
\affiliation{ Dipartimento di Fisica,   Universit\`a di Pisa, largo B. Pontecorvo 3, I-56127 Pisa, Italy,}

\author{Angela D.V. Di Virgilio}
\affiliation{Istituto Nazionale di Fisica Nucleare (INFN), sez. di Pisa, largo B. Pontecorvo 3, I-56127 Pisa, Italy.}

\date{\today}

\begin{abstract}

The study of local deformations is a hot topic in geodesy. Local rotations of the crust around the vertical axis can be caused by deformations. In the Gran Sasso area the ring laser gyroscope GINGERINO and the GNSS array are operative. One year of data of GINGERINO is compared with the ones from the GNSS stations, homogeneously selected around the position of GINGERINO, aiming at looking for rotational signals with period of days common to both systems. At that purpose the rotational component of the area circumscribed by the GNSS stations has been evaluated and compared with the GINGERINO data.
The coherences between the signals show structures that even exceed 60$\%$ coherence over the 6-60 days period; this unprecedented analysis is validated by two different methods that evaluate the local rotation using the GNSS stations. The analysis reveals that the shared rotational signal's amplitude in both instruments is approximately $10^{-13} rad/s$, an order of magnitude lower than the amplitudes of the signals examined using the coherence method.
The comparison of the ring laser data with GNSS antennas provides evidence of the validity of the ring laser data for very low frequency investigation, essential for fundamental physics test.

\end{abstract}


\maketitle


\section{\label{sec:level1}INTRODUCTION}
The GINGER (Gyroscopes IN GEneral Relativity) project is aiming at building an array of ring laser gyroscopes (RLGs) in the underground Gran Sasso laboratory ("Laboratori Nazionali del Gran Sasso", LNGS). RLGs are top sensitivity devices able to measure angular rotation rate exploiting the Sagnac effect. GINGER will provide unique data for geophysical and fundamental physics investigation \cite{GINGER}, especially for very low frequency investigation. In this paper the data recorded by the RLG prototype GINGERINO \cite{GA5}, presently active at LNGS, are compared with the GNSS data in order to provide a test of the validity of the signal reconstruction of the RLG signals.
The GNSS technique is normally used for studies of the 3D kinematic deformation patterns on the surface of the Earth in local or global areas. In particular, GNSS time series allow to reconstruct plate kinematics at global scale (e.g. \cite{KREEMER14,ALTAMIMI23}) and local strain rate tensors that identify deformation zones at any scale on the Earth \cite{OKAZAKI21,SERPELLONI22,PALANO23}. Here we exploit the possibility that a GNSS network and the Ring Laser Gyroscope (RLG) facility may both sense a regional rotational signal of the ground motion. Recent publications demonstrate that ground deformations in seismically active areas are not only dominated by tectonic motions but are also influenced by important hydrological processes \cite{LEONE23,PINTORI21,MICHEL21,DEVOTI15}, non-tidal and atmospheric loading processes \cite{MEMIN20,MARTENS20}. All these processes may, in principle, induce regional deformations with periodicity ranging from a few days up to multi-annual signals, mainly driven by climatic variations of the water cycle.\\
In this low frequency range the use of RLG data may be a promising tool to monitor local geodetic signals showing their rotational nature.
GINGERINO is sensitive to rotations around the vertical axis of global and local origin \cite{diciotto}. It is the first RLG based on a simple mechanical structure which has been able to run in a continuous basis with duty cycle larger than $90\%$.\\
The aim in this work is to define a relevant set of observable that can be used to correlate GNSS and RLG observations, in order to isolate possible regional geophysical processes that induce significant rotation rates. To our knowledge this is the first attempt to observe common signals in these instruments. \\
In the following the first section is dedicated to the description of the GINGERINO apparatus and the analysis applied to the data to be compared to the GNSS array. In the second, we illustrate the methods used to derive the rotational component of GNSS stations. Next, we show some tests performed on the analysis tool used to derive coherences and in the last section some comments on the results and how to extend the analysis in the future. 

\section{GINGERINO, the RLG prototype of the Gran Sasso}

\begin{figure}
    \centering
    \includegraphics[scale=0.26]{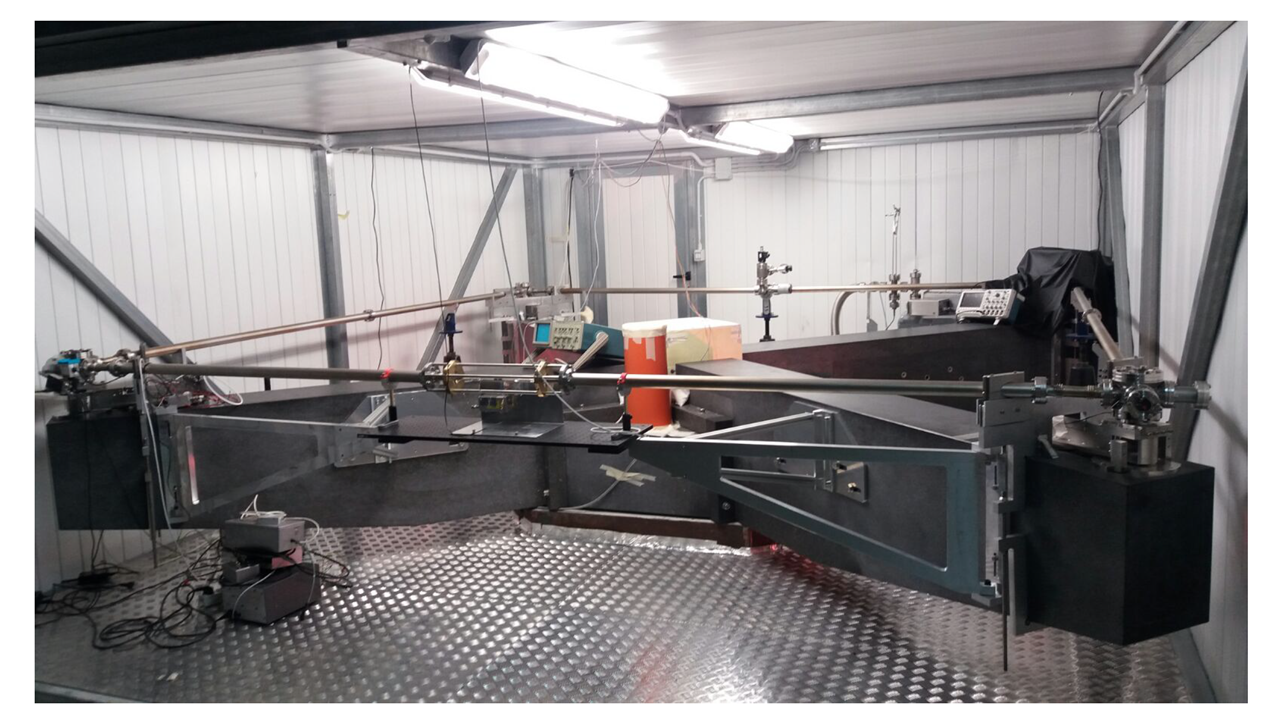}
    \caption{GINGERINO is composed of a cross shaped granite monument, connected to the floor through a reinforced concrete monument. 
    On top of the granite cross, four boxes are attached, where the mirrors are located.
    The four boxes are connected by pipes where the light travels. In first plane, in between two mirror boxes the discharge tube is visible, necessary to start and keep working the laser. }
    \label{fig:GING}
\end{figure}

The Sagnac effect states that two counter-propagating light beams traveling in a closed ring path complete the round trip with a difference in time proportional to the absolute (or inertial) angular rotation rate of the rigid closed path itself.\\
Based on this principle several devices have been developed for inertial navigation, using optical fibers, optical cavity and cold atoms. A RLG is composed of a high finesse polygonal optical cavity, where an active medium is contained. Exciting the active medium the laser process is generated in the two directions. If the optical cavity is rotating, the frequency of the two counter-rotating beams is different, due to the different round trip time. Then, combining outside the cavity the two beams  on a photodetector, a beat signal is generated at the so called Sagnac frequency$f_s$:

\begin{equation}
    f_s = 4 \frac{A}{P\lambda}\cdot \Omega \cos{\theta}
    \label{eq:uno}
\end{equation}

where A is the area delimited by the light path, P the perimeter, $\Omega$ the absolute value of the angular velocity of the optical cavity, $\lambda$ the laser wavelength and $\theta$ is the relative angle between the direction (or rotation axis) of $\vec{\Omega}$ and the area vector of the cavity $\vec{A}$. GINGERINO, see Fig. \ref{fig:GING} is rigidly attached to the ground, has square optical cavity with 3.6m side, lying on the horizontal plane (vertical $\vec{A}$), so the angular velocity is mainly $\Omega_{\oplus}$, the Earth rotation rate, $\theta$ is the co-latitude. The  sensitivity has been already demonstrated to be in the range of pico-rad/s, with detection bandwidth of tens of Hz. 
In the comparison with GNSS signals the GINGERINO data has been processed in order to remove environment, tides and laser disturbances, as explained in the following subsection.

\subsection*{Details of the procedure for GINGERINO signals reconstruction}

The radiation of the two counter-propagating beams transmitted  by one mirror are recombined and their interference is detected on a photodiode.
GINGERINO is connected to the Earth crust and the Earth rotation $\Omega_{\oplus}$ is the largest signal, the interference is the beat note, $ \omega_m $. \\
It is important to note that  $\omega_m$ cannot be confused with  the Sagnac frequency, $\omega_s = 2 \pi f_s$, since the interference signal contains small perturbations induced by the diffused radiation by the cavity mirrors and the laser non linear dynamic. It is necessary to reconstruct $f_s$ using the signals exiting the RLG \cite{dieci,AA26}: the beat note and the intensities of the two counter-propagating light beams (mono-beams). In the present analysis the procedure to evaluate $f_s$ follows what described in previous papers \cite{PRR2020, EPJC2020, EPJC2021}, where more details can be found about the theory of analysis procedure and how the terms necessary to eliminate the typical laser disturbances called back--scatter and null shift are evaluated and used. It is important to say that those laser systematics affect $\omega_m$ mainly below 1Hz.\\
The optical cavity of GINGERINO is free running, and because of temperature variations, the perimeter of the optical cavity changes, and as a consequence, also the laser emitted wavelength changes. Accordingly, sometimes the light beams change abruptly its frequency to go back in a position with higher gain. Two possible effects take place: mode jumps and split mode operations. 
In the first one the two counter--propagating beams change optical frequency at the same time, a very fast transient happens with a few seconds duration. In the second one only one beam changes frequency and the two beams have difference in frequency of the free spectral range of the cavity, so that $\omega_m$ has very high frequency (MHz), and the GINGERINO apparatus is not able to acquire. Split mode operation can last for several hours and the data are lost, but are rare. The data are selected eliminating the portions with split mode operation or affected by mode jumps, and others short duration large signals, in the present analysis $20\%$ of the data are eliminated. GINGERINO is isolated and based on a rather rigid structure, it is assumed that external variations and signals will affect $\omega_m$ at first order, accordingly the different expected signals and environmental disturbances are estimates and subtracted with statistical means, basically using standard linear regression. For the linear regression (LR) it is necessary to elaborate a set of terms to be used in the linear regression. For the GINGERINO analysis the following terms are used: 
\begin{itemize}
\item the laser dynamic has to be subtracted, so several terms are elaborated following the theory (called LD)
    \item the polar motions, called $F_{geo}$, the angular rotation around the vertical axis as expected at the GINGERINO latitude and longitude estimated by the international system IERS, 
    \item tides affect the apparatus, and the full list of tides elaborated using the program  $GOTIC2\_{mod}$ , are used as LR terms and called $T$,
    \item GINGERINO is a mechanical apparatus and accordingly temperature  causes geometry changes, deformation, orientation and  the environmental monitors (EM) of the GINGERINO apparatus, as tilts, temperature and pressure are used as terms of the LR.
\end{itemize} 

The first step of analysis provides a quantity $\omega_{s0}$, which well reproduce the true Sagnac frequency above $1mHz$, and evaluates all the required terms for the LR: the $LD$ (terms to subtract the Laser dynamic), $F_{geo}$, $T$ and $EM$. 
The LR procedure is iterative, using each iteration 3 consecutive days, and keeping only the central day, so the coefficients of the linear regressions changes day by day, but are evaluated using 3 days. It is important to remind that the linear regression is done proposing about 10 terms, and  selecting the ones with p-value below 0.1.

\begin{figure}
    \centering
    \includegraphics[scale=0.32]{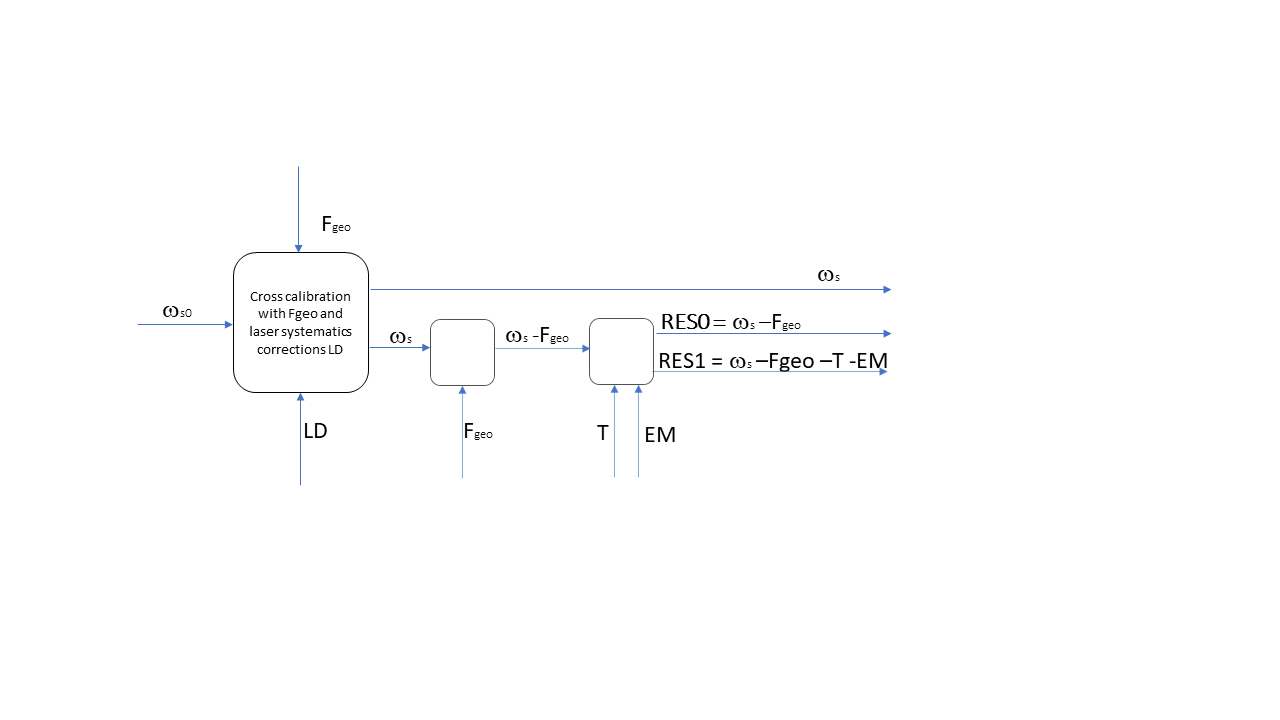}
    \caption{General scheme of the analysis procedure.}
    \label{fig:layout}
\end{figure}

The full analysis releases three different outputs, see Fig. \ref{fig:layout},  the first level uses $F_{geo}$ and $LD$, minimising the quantity $F1$ defined as: 
\begin{equation}
    F1 = \omega_{s0} - k\cdot F_{geo} - \mu\cdot LD 
\end{equation}
 $k$, and $\mu$ are the coefficients found by the linear regression to minimize $F1$, and 
 \begin{equation}
\omega_s = F1/k +F_{geo}
 \end{equation}
Calling $\omega_s$ the true angular rotation recorded by GINGERINO, the term $k$ mitigates for the 3 days  used in the LR effects due to scale factor variations due to temperature variation and change of inclination of the RLG optical cavity.
The first residuals $RES0$ represent the true angular rotations cleaned by the polar motion effect: 
\begin{equation}
    RES0 = \omega_s - F_{geo}
\end{equation}

Last step $RES1$ minimises the residuals subtracting $T$ and $EM$:
\begin{equation}
     RES1 = RES0 - b\cdot T - c\cdot EM
\end{equation}
$b$ and $c$ are the coefficients of the linear regression.

The three quantities $\omega_s$, $RES0$ and $RES1$ are interpolated in order to fill the gaps due to mode jumps and split modes and used in the comparison with GNSS data.

\section{Coherence GINGERINO-GNSS stations signal }

\begin{figure}[h]
    \centering
    \includegraphics[width=0.5\textwidth]{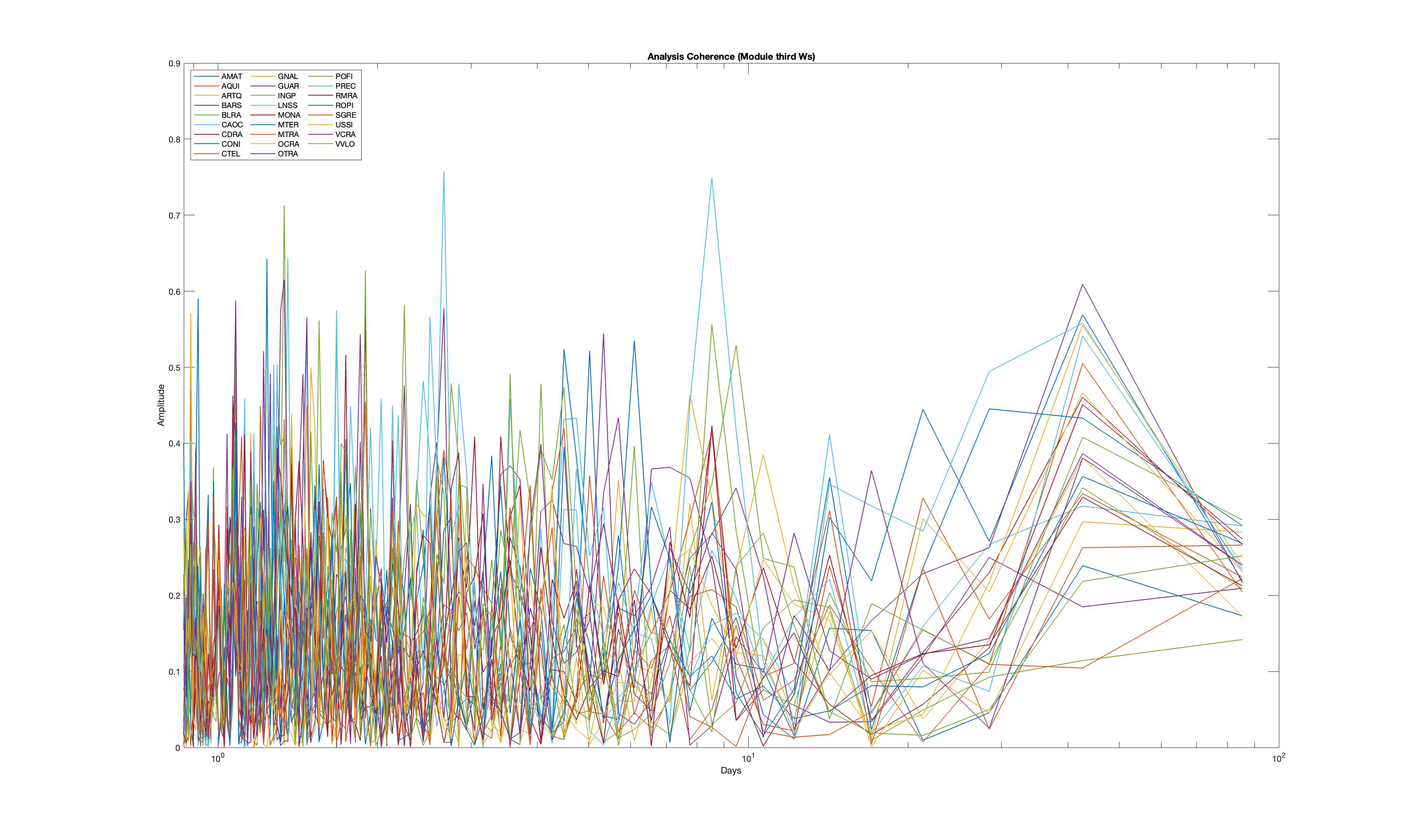}
    \caption{Coherence between individual stations and GINGERINO}
    \label{fig:Coh1}
\end{figure}

\begin{figure}[h]
    \centering
    \includegraphics[width=0.55\textwidth]{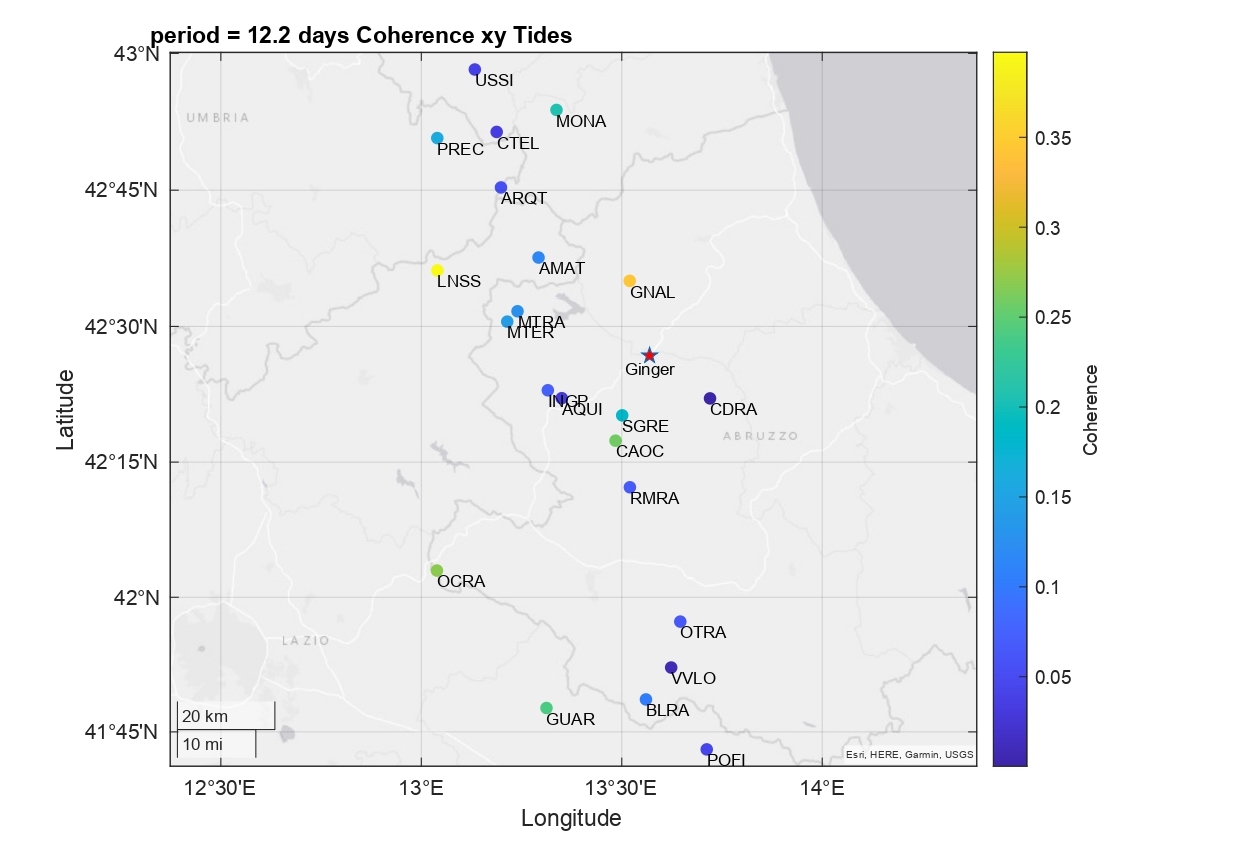}
    \includegraphics[width=0.55\textwidth]{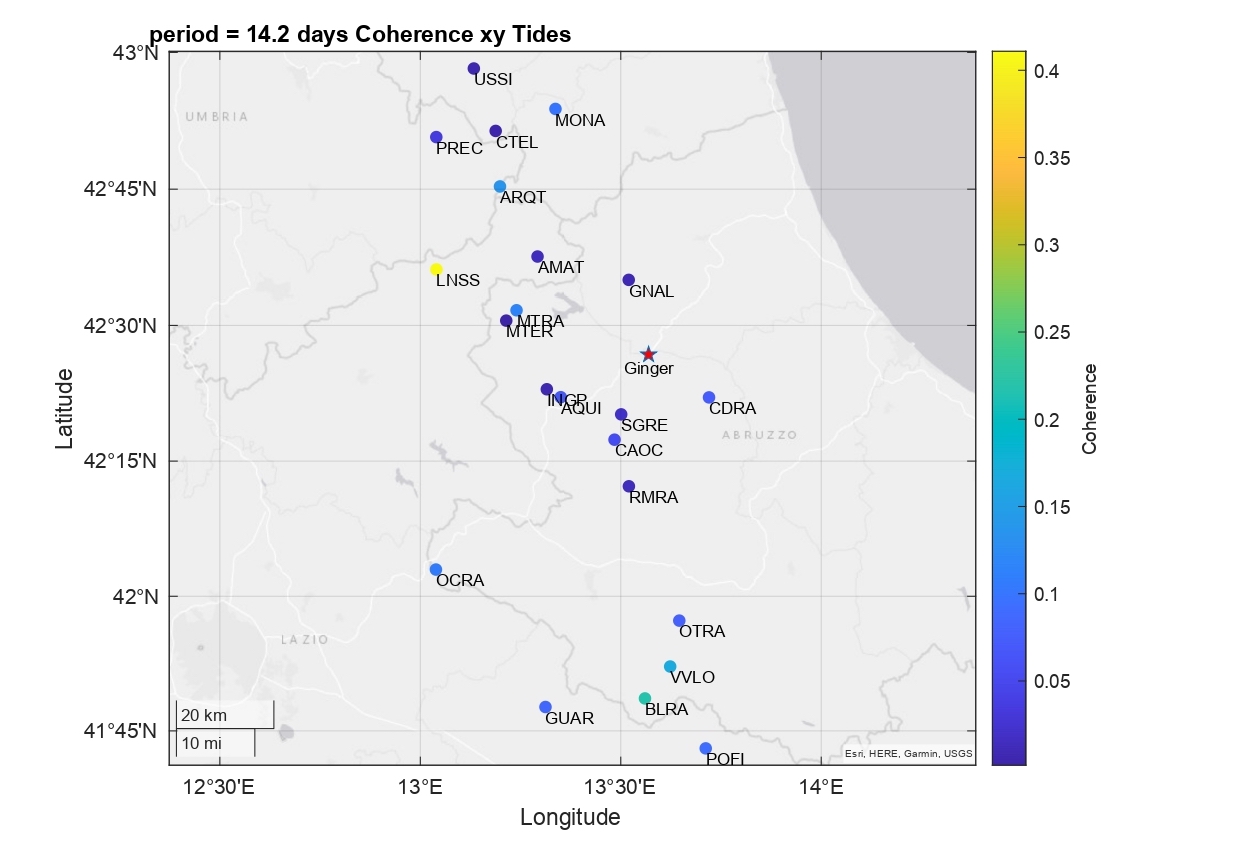}
    \caption{The two plots show the topographic trend of coherence at the choice of specific frequencies. The GINGERINO signal used is the one in which the tides were resolved.}
    \label{fig:Coh2}
\end{figure}

\begin{figure}[ht]
    \centering
    \includegraphics[width=0.55\textwidth]{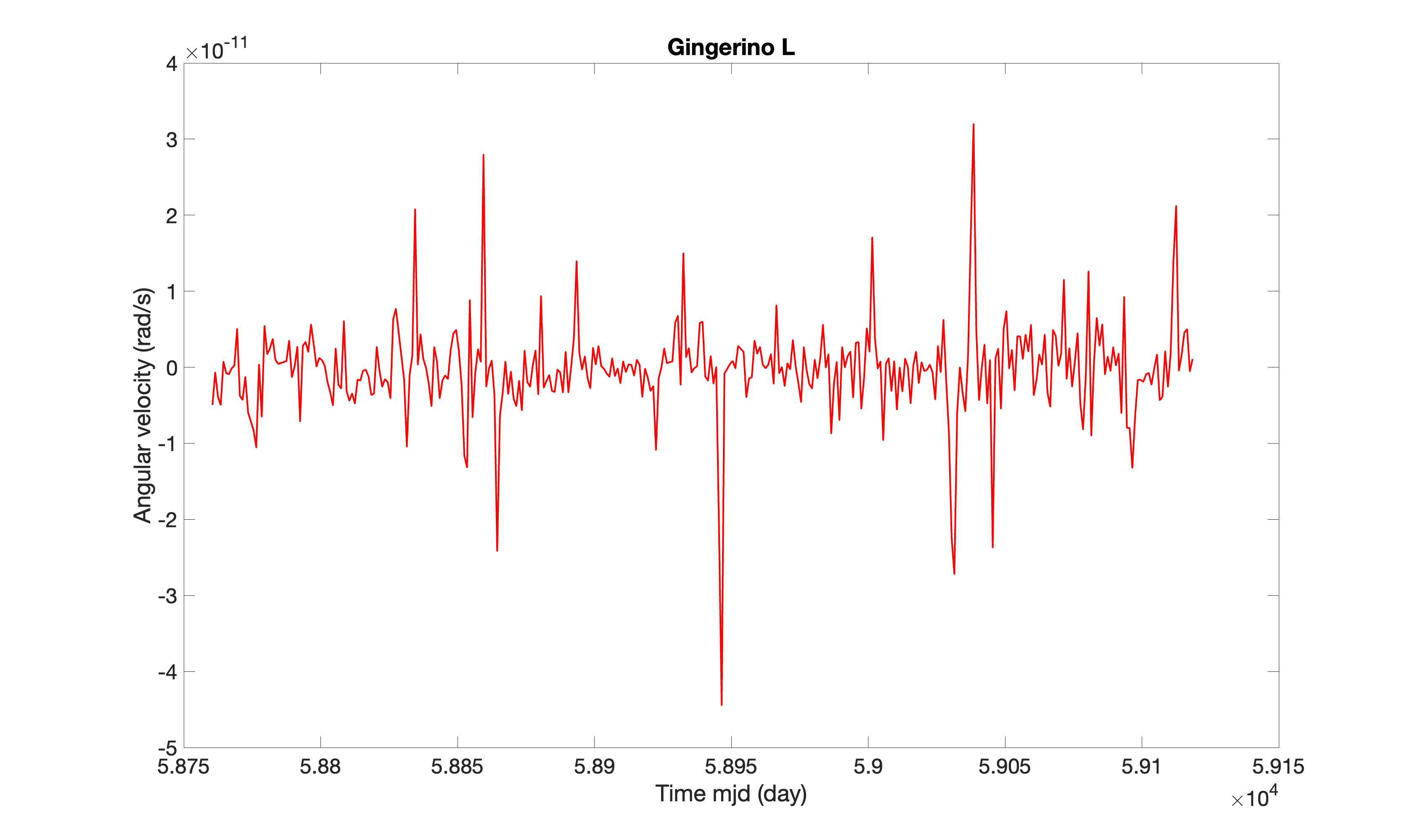}
    \includegraphics[width=0.55\textwidth]{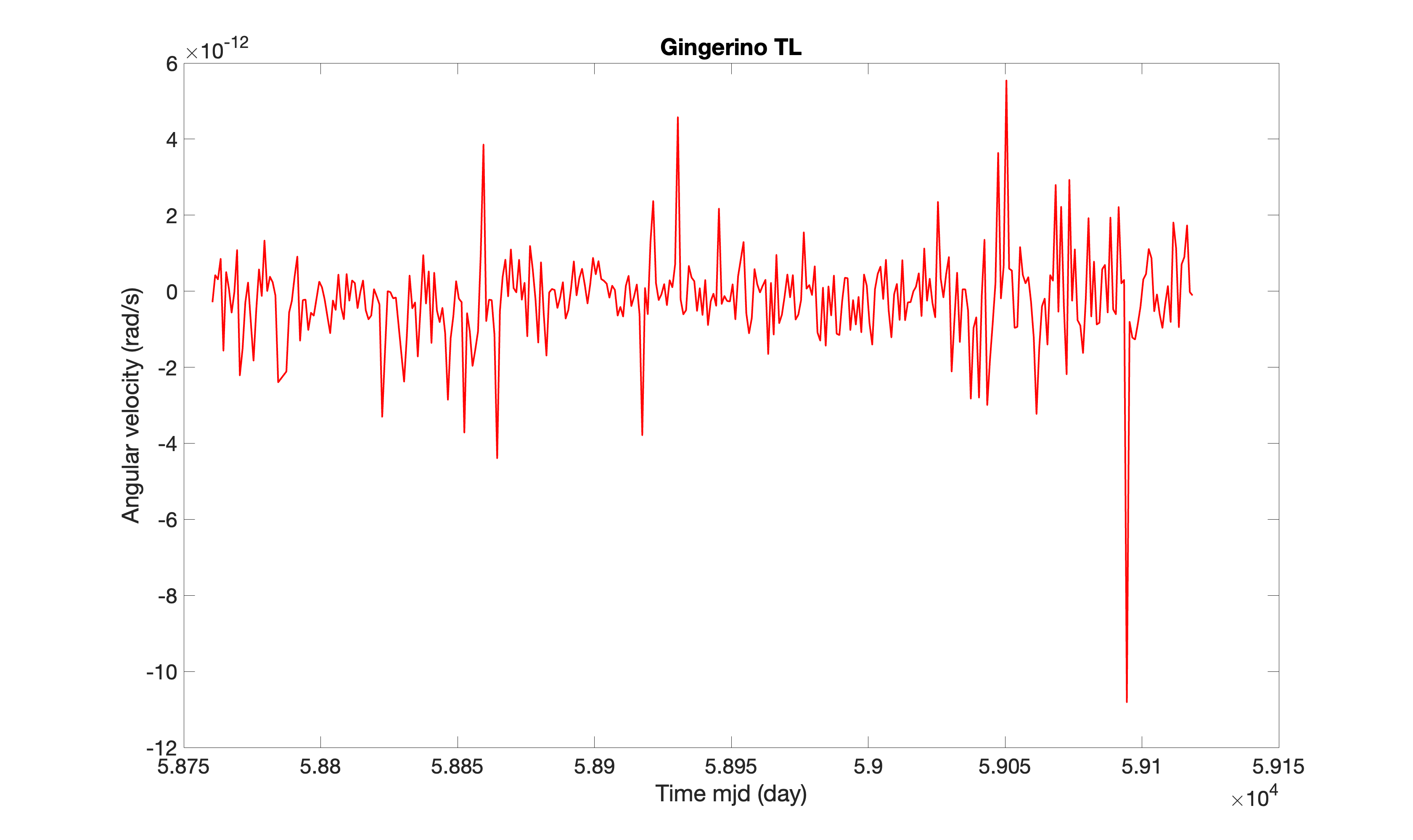}
    \caption{At the top the GINGERINO signal in which systematic laser corrections and terrestrial rotational component, including polar motion and Chandler wobble (obtained from IERS measurements \cite{IERS}), were removed.
At the bottom we have the GINGERINO signal, obtained starting from the previous one, in which we solved and subtracted the tides through the use of the $GOTIC2\_{mod}$ program \cite{GOTIC2}. }
    \label{fig:Coh3}
\end{figure}

To derive a topographic trend over a period of 2 to 256 days, we initially calculate the coherence between GINGERINO and individual stations. However, upon examining the plot in Fig.\ref{fig:Coh1}, which considers the contribution of individual stations, no apparent common structures or trends are evident. Consequently, we narrowed our focus to periods of 12-14 days and around 7 days of interest, regions were typically tides and signals related to anthropic activity are present. \\The results in Figs. \ref{fig:Coh2},\ref{fig:Coh3} demonstrate that there is no discernible spatial pattern either by moving away from GINGERINO's position or by considering different areas to the east and west of the Apennines. Additionally, the coherence value falls short of 40$\%$, a comparatively low value. As we will delve into later, it's worth noting that even two random noise signals can attain a coherence value of 37$\%$ using the same analytical tools at our disposal.
\\

\section{ Evaluate rotations using the GNSS stations } 

As previously explained, GINGERINO measures angular rotation rate around its vertical axis, while the GNSS stations measures 3D displacements. To make a step forward it is necessary to evaluate the rotation using the GNSS array. For this purpose we employed two different technique.\\
The crust exhibits both strain and rotational deformation components, implying that individual stations are spatially correlated.
In the subsequent sections, the two methods employed to determine the rotational component are described. Firstly, we represent it as the straightforward rotational velocity around a vertical axis. Secondly, we derive it as anti-symmetric rotational component of the strain-rate matrix, which can be effortlessly traced back to the rotor located at the GINGERINO position. One year of data of GINGERINO is compared with the ones from the GNSS stations. When accompanied by "$mjd$", the time is expressed in the modified  Julian date.

\subsection*{Rotational components approach}

\begin{figure}[h]
    \centering
    \includegraphics[width=0.53\textwidth]{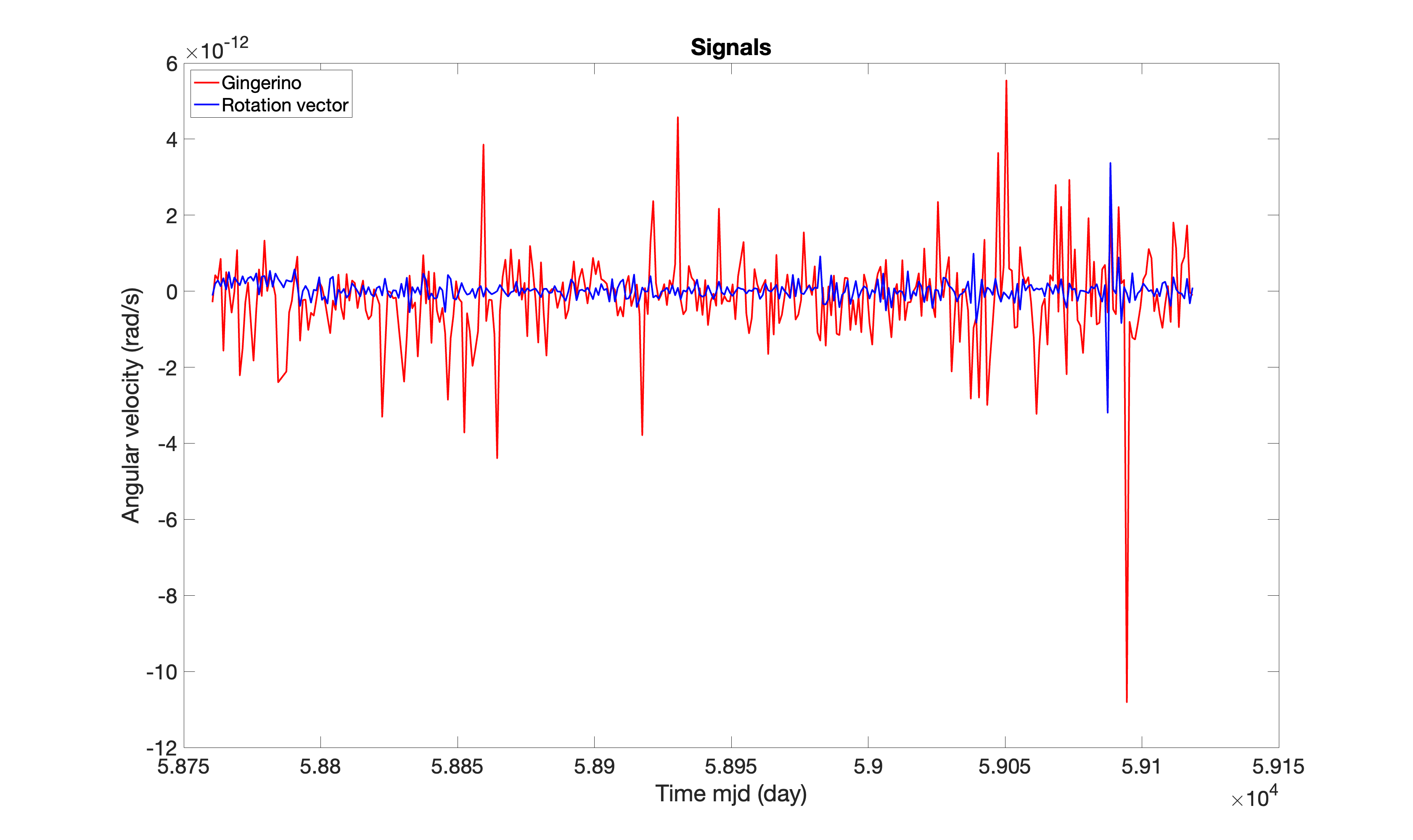}
    \caption{Shown in red is the tide-cleaned GINGERINO signal and in blue that obtained by the rotation vector method, the period under examination is 359 days.}
    \label{fig:DiSP}
\end{figure}

The first method to derive the rotational component from a network of GNSS stations is to directly calculate the rotation vector associated with each individual station and sum them appropriately; in our specific case, for reasons mentioned above, the chosen rotation pole is the position of GINGERINO.
Using eq. \ref{w1} we can calculate the magnitude of the angular velocity vector associated with each GNSS station:

\begin{equation}
    \omega_1 = \frac{|v_1|}{|r_1|}\sin{(\alpha_1-\theta_1)} 
    \label{w1}
\end{equation}
where $|v_1|$ is the magnitude of the horizontal velocity of the individual station seen in the Terrestrial Reference System integral to the Erthe's crust  in which we decompose it into components of latitude $v_N$ and longitude $v_E$ as:  $|v_1|=\sqrt{v_E^2+v_N^2}$, 
while $\alpha_1$ is the angle that the velocity vector forms with the local parallel, that is: $
    \alpha_1=\arctan{\left(\frac{v_N}{v_E}\right)}
    \label{alpha1}$. 
Instead, $ |r_1|$, the absolute value of the station's distance from GINGERINO and $\theta_1$, the angle that the $\hat{r}_1$ vector forms with the parallel. They are derived with the Matlab function 'distance', specifying the coordinates of GINGERINO, those of the station and the Earth model, which in our case is the ellipsoid.
In order to evaluate the mean value of the rotational velocities of the stations array, the weights and thus the errors must be calculated; details on the error estimation can be found in the Appendix A. 
We present the signal obtained through this method in Fig. \ref{fig:DiSP}, wherein it is worth noting that despite having a smaller standard deviation, the signal exhibits amplitudes of comparable orders to that of GINGERINO and even some similar spikes. However, to ascertain whether both signals capture the same phenomenon, a coherence analysis is essential.

\subsection*{Curl approach}

A second method to derive the rotational component of the area circumscribed by the GNSS constellations is to calculate the z-component of the rotor at the GINGERINO position. To do this, we considered the equations linking the velocities of the stations $v_i$ with their positions $x_i$ \cite{Allmendinger}:

\begin{equation}
    v_i = t_i + \frac{\partial v_i}{\partial x_j} x_j = t_i + e_{ij} x_j
    \label{vel}
\end{equation}

where $t_i$ is a simple integration constant that takes into account the displacement at the origin of the coordinate system. The matrix $e_{ij}$ is the displacement rate gradient tensor, and through the:

\begin{equation}
    e_{ij} = \epsilon_{ij}+\omega_{ij}=\frac{(e_{ij}+e_{ji})}{2} + \frac{(e_{ij}-e_{ji})}{2} 
    \label{e}
\end{equation}

we can divide it into the symmetrical part the infinitesimal strain rate tensor and the anti-symmetrical part the rotation rate tensor. The latter, apart from a factor of 2, corresponds to the z-component of the curl:

\begin{equation}
    \omega_z= \left(\frac{\partial v_x}{\partial y}-\frac{\partial v_y}{\partial x}\right)
    \label{wz}
\end{equation}

\begin{figure}[ht]
    \centering
    \includegraphics[width=0.53\textwidth]{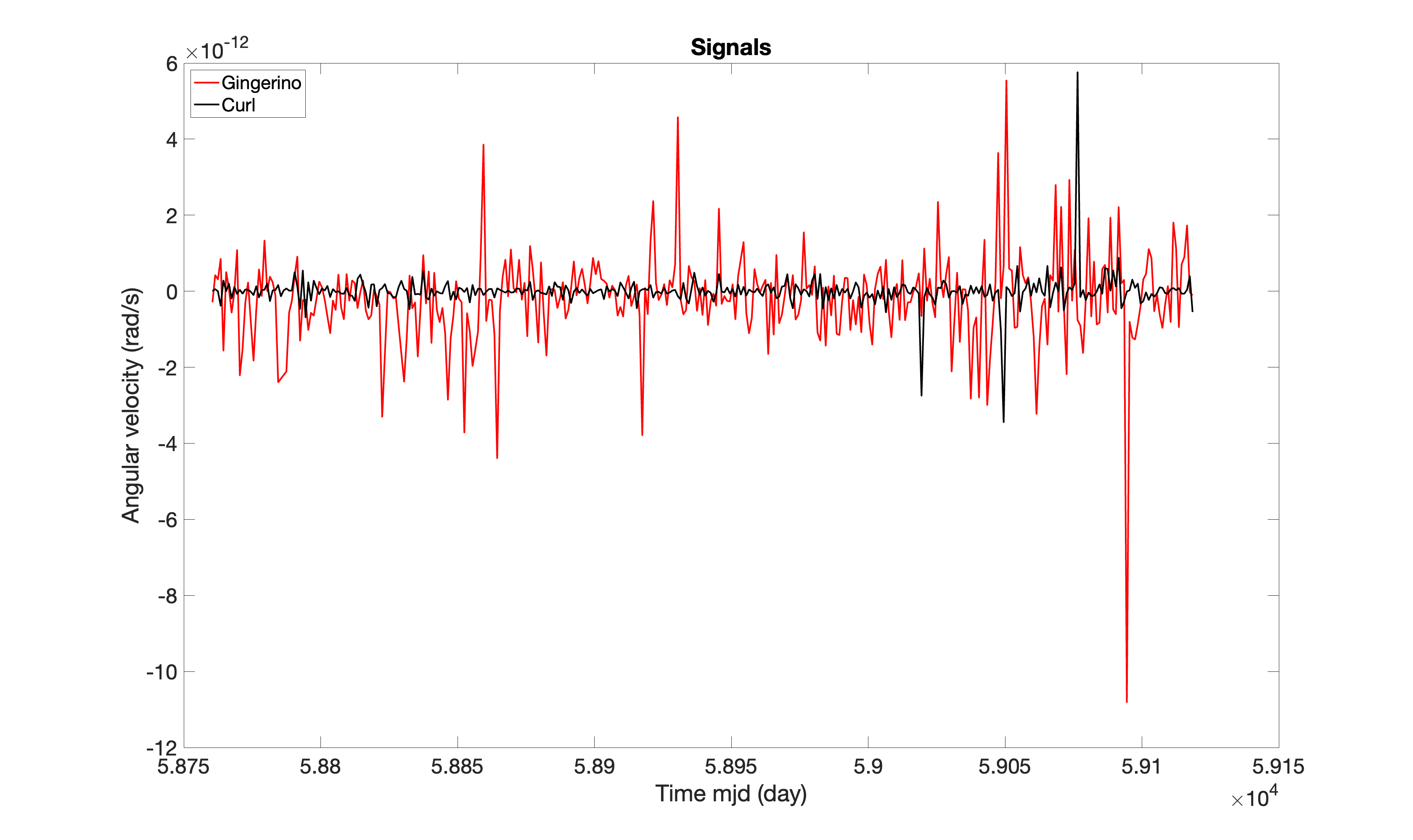}
    \caption{Shown in red is the tide-cleaned GINGERINO signal and in black that obtained by the Curl method, the period under examination is the same as the previous one.}
    \label{fig:DevErr}
\end{figure}

\begin{figure}[ht]
    \centering
    \includegraphics[width=0.53\textwidth]{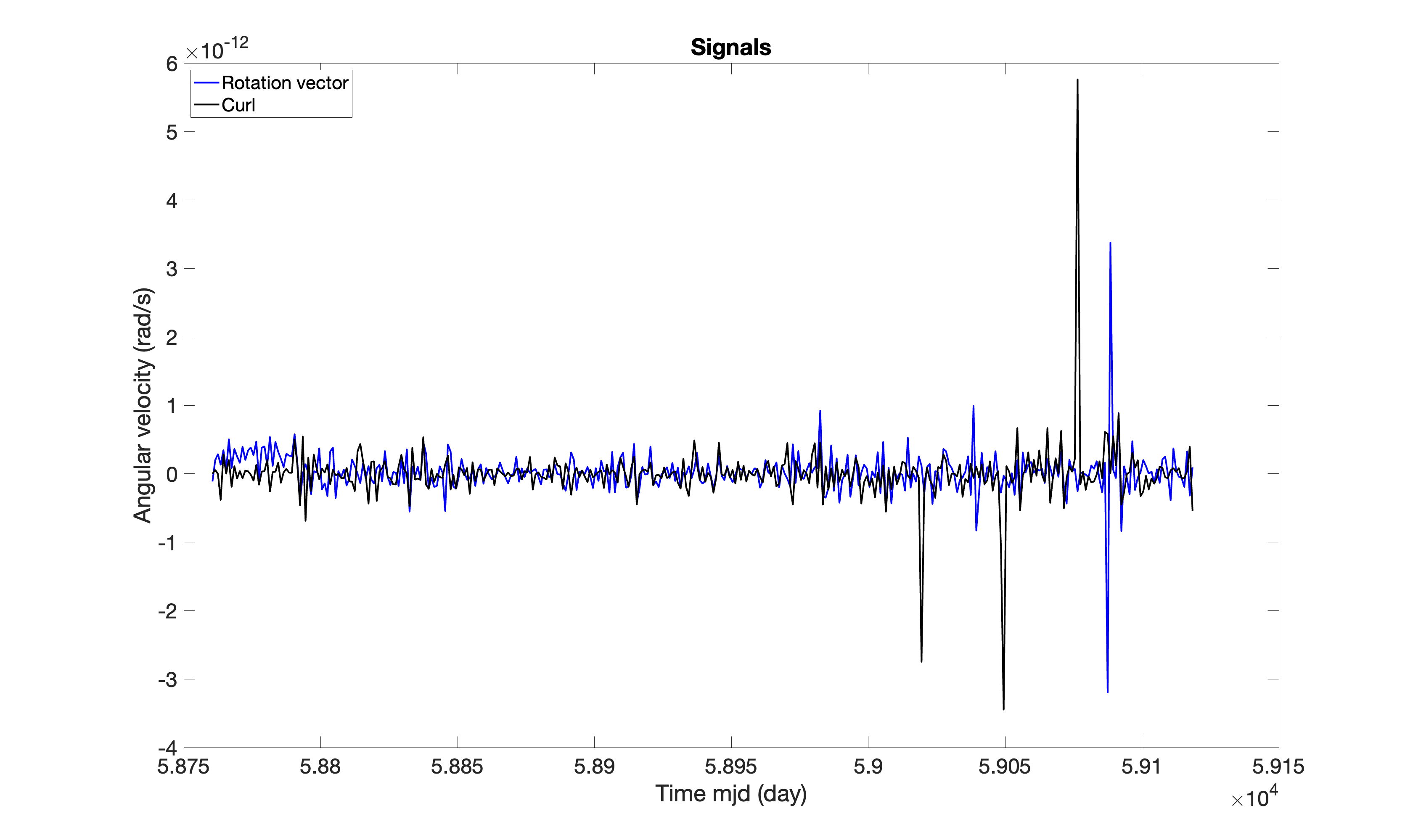}
    \caption{Comparison of the two method. }
    \label{fig:CurlRot}
\end{figure}

Solving these equations involves a well-known inverse problem. To tackle this, we employ a conventional least squares approach that effectively handles the overconstrained nature of the problem. \\
To do this, we minimise the expression:

\begin{equation}
    \left[\overline{v}' - A \cdot \overline{x}'\right]^2
    \label{Minim}
\end{equation}

where the matrix $A$, called the design matrix, takes into account how the velocity components of the individual stations have been arranged, it is the matrix transposition of the equations \ref{vel}.\\
As previously stated, the crust supporting the stations adheres to the principles of quasi-rigidity. While the stations are free to move, but they display interconnections and correlations. Unlike the previous approach, we compute the rotation values by incorporating the signals from the stations simultaneously. To adequately address the correlation between these signals, we construct a block matrix consisting of sub-matrices in a diagonal format:

\begin{equation}
    Corr =  \begin{pmatrix}
 \sigma_x^2 & C_{xy} & C_{xz}\\
 C_{xy} & \sigma_y^2 & C_{yz}\\
 C_{xz} & C_{yz} & \sigma_z^2
 \end{pmatrix}
 \label{CovVel}
\end{equation}

where we have $C_{ij} = R_{ij}\sigma_i\sigma_j$, both the standard deviations $\sigma_i$ and the correlation coefficients $R_{ij}$ are obtained directly from the data provided by the GNSS stations, for example, in our case, having used 22 GNSS stations we obtained a block matrix 66x66 and each block on the diagonal corresponds to a single station with a 3x3 matrix given by eq. \ref{CovVel}.
Let us show in Fig. \ref{fig:DevErr} the comparison between the signal of GINGERINO and the one obtained with the Curl method, we note how it is also for the previous method that the magnitude of the signal amplitudes is the same for both GINGERINO and the methods just right ($10^{-12} rad/s$), this is surprising since, as can be seen from the top part of the Fig. \ref{fig:Coh3}, if we had not resolved the tides for GINGERINO, as they are resolved in GNSS stations, we would not have obtained the same magnitude but GINGERINO would be an order of magnitude above. We show the direct comparison between the two signals in Fig.\ref{fig:CurlRot}.

\section{Coherence analysis}

Before delving into the analysis of the correlations between the two methods and the GINGERINO signal, let's conduct some tests using the MATLAB function "mscohere." These tests aim to determine the level of confidence required to validate a coherence signal. Additionally, we'll explore how low we can go below the order of the signals under consideration to find a positive coherence signal.
Lastly, we will perform a fitting procedure on the station velocities in an attempt to enhance the signal-to-noise ratio for the implemented methods.

\subsection*{Baseline for zero coherence: ''mscohere''}

\begin{figure}[h]
    \centering
    \includegraphics[width=0.53\textwidth]{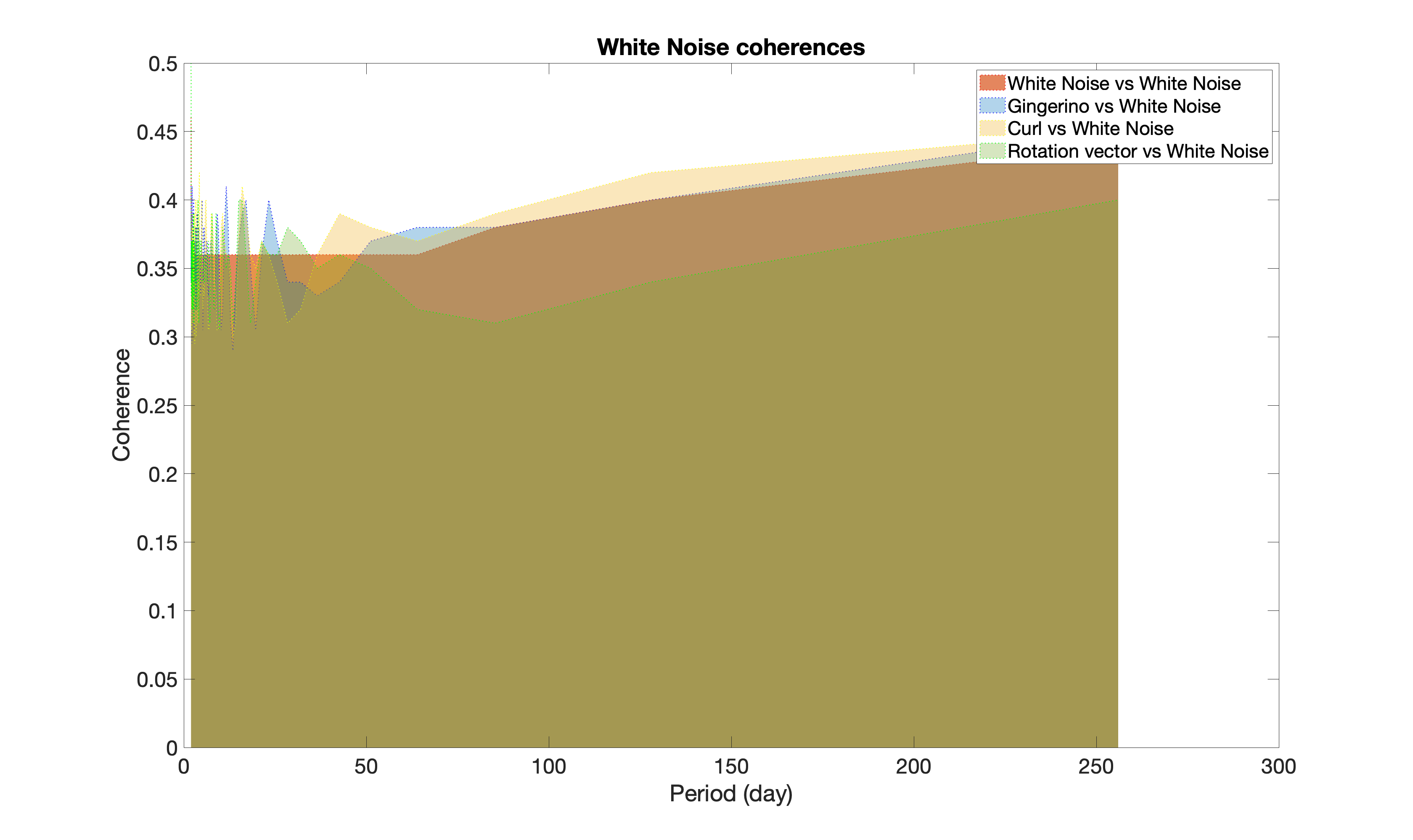}
    \caption{We show here the coherence with 95$\%$ confidence intervals for all simulations made in the comparison with synthetic random noise.}
    \label{fig:Coh4a}
\end{figure}

To determine the actual degree of coherence between the two signals, we conducted tests using the ''mscohere'' function along with simulated random noises. Employing a Monte-Carlo simulation approach, we generated 100,000 instances of random noise, using the function $rand$ \footnotemark, and calculated the coherence with respect to: another random noise, the Curl signal, the rotation vector signal, and the GINGERINO signal. For each of the aforementioned comparisons, we calculated the upper limit at a 95$\%$ confidence level for every frequency. The resulting data is presented in Fig. \ref{fig:Coh4a}. Notably, despite considering the comparison with random noise, we observed non-zero coherence levels. In fact, for the majority of ports, the frequency band reached 37$\%$, furthermore, by choosing this value as zero we have a conservative approach, however, it remains uncertain whether structures below this confidence level are real or not. Hence, we establish this value as the new baseline for zero coherence, above which we conclude that two signals exhibit coherence associated with a physical signal; in addition, the simulation between two random noises has a linear edge, while small structures appear between the rotation and random noise signals.
\footnotetext{$rand$ generates uniformly distributed random numbers in the interval [0, 1), but the same result is achieved by using the $randn$ function, which a instead generates random numbers distributed according to a standard normal distribution.}

\subsection*{Detection of a synthetic signal at a known frequency}

\begin{figure}[ht]
    \centering
    \includegraphics[width=0.51\textwidth]{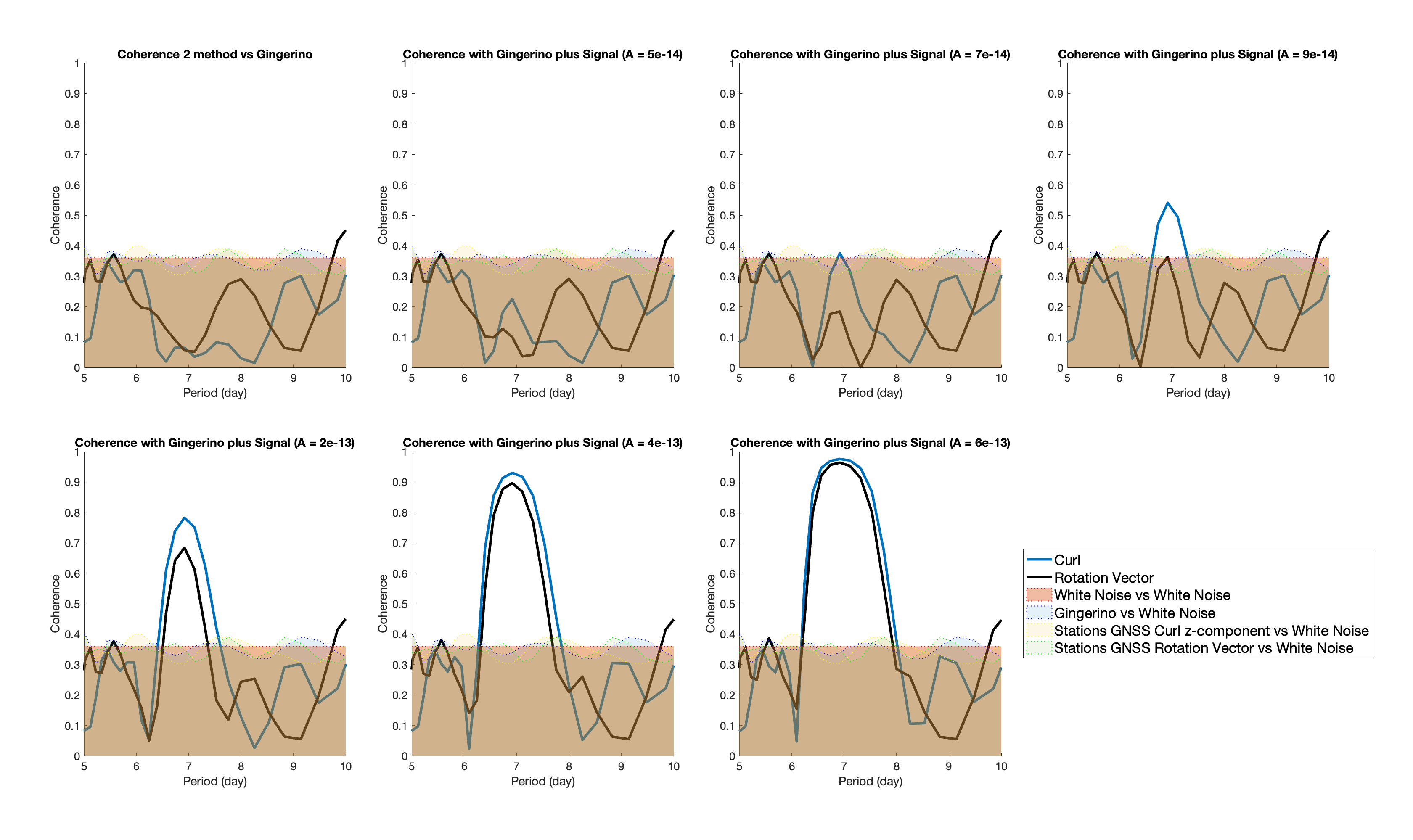}
    \caption{In the first quadrant, we show the coherence between GINGERINO and the GNSS data, using the two analysis methods over an interval of 5 to 10 days. In the following quadrants we show by how this coherence varies by adding to the signal a sinusoidal signal, that exhibited spikes over a duration of 7 days, with amplitudes: $5\cdot10^{-14}rad/s$, $7\cdot10^{-14}rad/s$, $9\cdot10^{-14}rad/s$, $2\cdot10^{-13}rad/s$, $4\cdot10^{-13}rad/s$, $6\cdot10^{-13}rad/s$. }
    \label{fig:CohTest7}
\end{figure}

We validated the analysis method by injecting a synthetic signal into all the signals we considered. The objective was to assess the effectiveness of the coherence method in producing positive results when applied to the signals under analysis. Specifically, we introduced a sinusoidal signal with a variable amplitude and a fixed frequency corresponding to a 7-day period. We selected this specific period because the coherence method does not exhibit any significant signals within that frequency range.
By referring to Fig. \ref{fig:CohTest7}, we observe that the synthetic signal remains below the noise level established in the previous test, as long as the amplitude does not exceed $9\cdot10^{-14}rad/s$. Considering that the signals under examination have an amplitude of $10^{-12}rad/s$, we can conclude that this analysis method allows us to identify genuine coherence peaks for signals that are approximately one order of magnitude lower than the maximum amplitude of the signal being analyzed.

\subsection*{Fit of linear velocities}

\begin{figure}[h]
    \centering
    \includegraphics[width=0.51\textwidth]{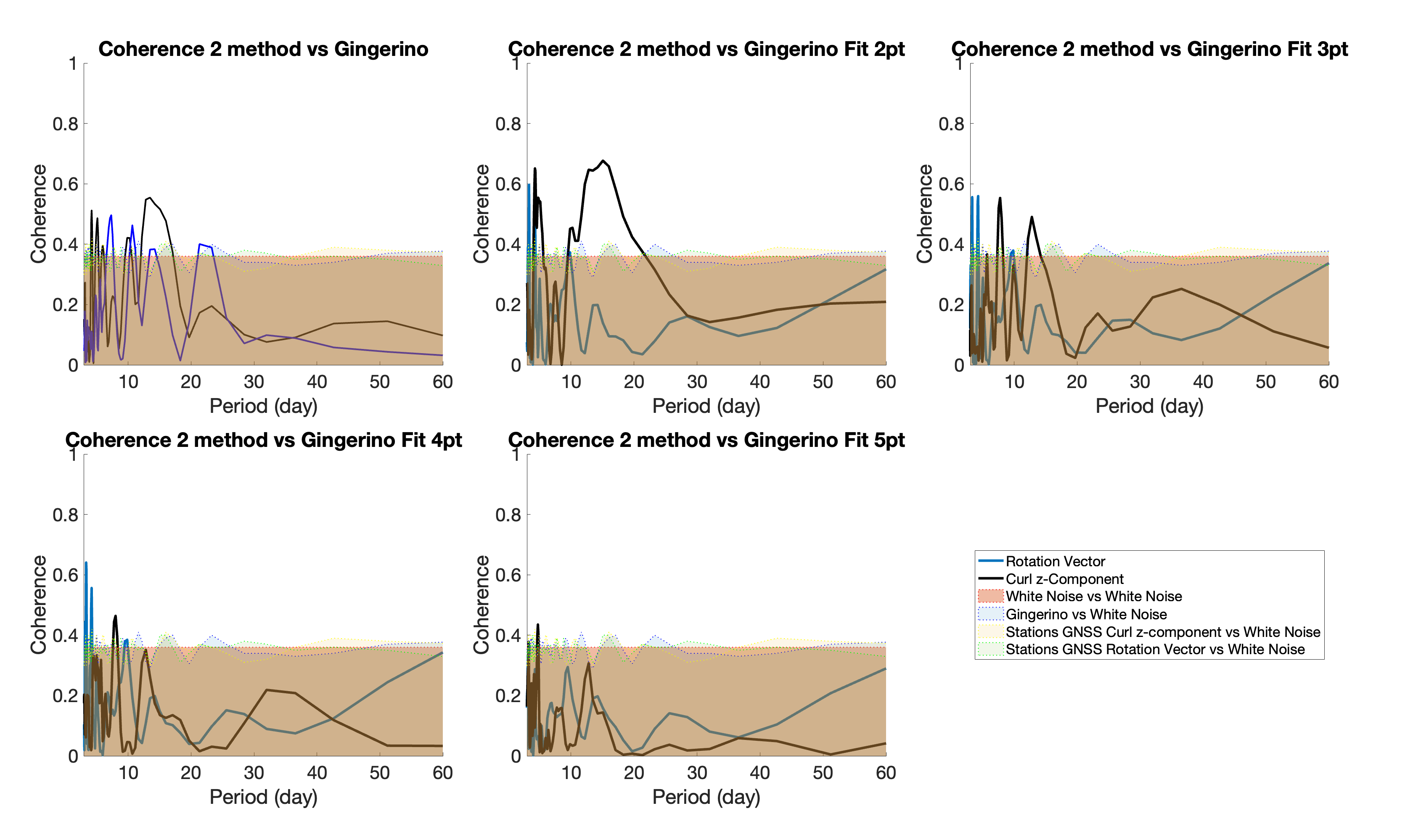}
    \caption{Coherence obtained without subtracting the contribution of the tides, clear structures are evident with the 2-point fit. }
    \label{fig:Coh4b}
\end{figure}

\begin{figure}[h]
    \centering
    \includegraphics[width=0.51\textwidth]{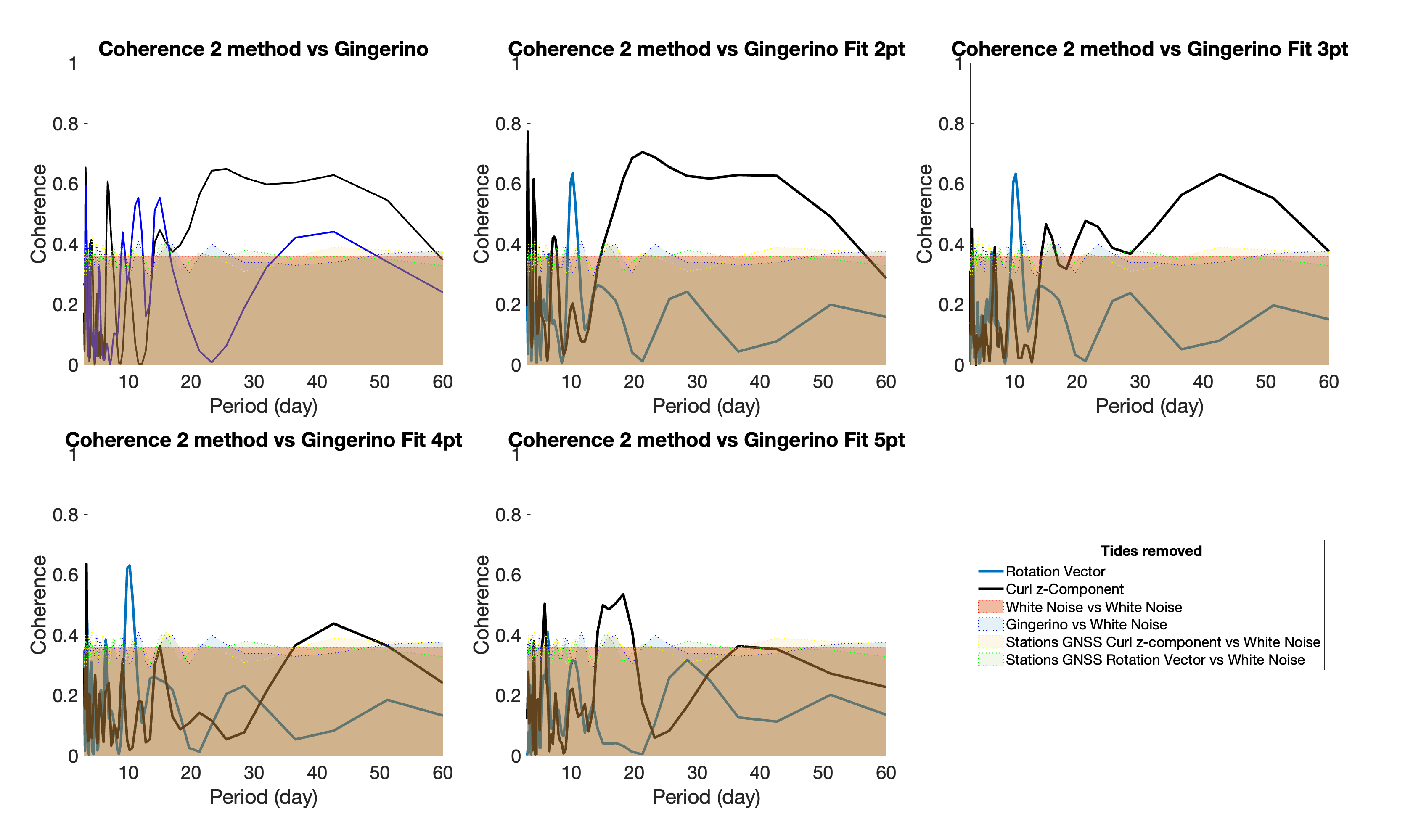}
    \caption{In this figure, we observe the coherence achieved through the resolution of tides in GINGERINO. The usual tidal peaks are reduced, revealing previously hidden structures with periods exceeding 20 days. }
    \label{fig:Coh4c}
\end{figure}

In the previous described analysis we have determined speeds of GNSS stations by comparing their different positions on consecutive days. The noise  can be reduced by a moving average approach by a linear fitting through multiple data points.\\
For this purpose, we utilize the "polyfit" function in Matlab. This function takes into account the positions of individual stations along with their corresponding errors, and outputs the angular coefficient and its associated error for the best-fit line. We conduct the analysis repeatedly, considering 2, 3, 4, and 5 points at a time. Each time, we shift the analysis by one day to maintain always the same number of points as the analysis without fitting.\\
The results of these analyses, incorporating the GINGERINO signal that includes tides, are presented in Fig. \ref{fig:Coh4b}. Additionally, we provide the results in Fig. \ref{fig:Coh4c}, where the tides have been resolved.

We observed that in the coherence analysis of the GINGERINO signal, which includes tidal effects, a prominent peak emerges at approximately 14 days. However, this peak vanishes when we perform the coherence analysis for all signals, taking into account resolved tides. It is likely that the presence of this peak is attributed to a frequency associated with the rotational dynamics of tides, which is still present in both the GNSS stations and GINGERINO.\\
Once the tides are properly resolved, less resolved structures become apparent, exhibiting a periodicity ranging from 20 to 60 days. These structures do not originate from tides. Interestingly, they are only visible when using the Curl method, possibly due to the distinct coordinate systems employed by the two methods. While the rotation vector (depicted in blue) solely considers the horizontal components, disregarding the vertical one, the Curl method utilizes geocentric coordinates, inherently encompassing the contribution from the radial component (vertical in the former method). Consequently, these poorly resolved structures, coming out of coherence noise, might be influenced by vertical phenomena impacting the Earth's surface. In the subsequent section, we will propose hypotheses to explicate these signals. \\
Additionally, it is worth noting that as we increase the number of data points used for the fitting process, the coherence level of these structures diminishes until it becomes indistinguishable from the previously calculated noise. This observation can be attributed to the decrease in signal amplitude resulting from the speed fit process. Consequently, the magnitudes of both the Curl and rotation vector signals become comparable to signals that are not discernible in coherence with the GINGERINO signal, which remains at a magnitude of approximately $10^{-12} rad/s$. This comparison renders these structures invisible in the coherence analysis, as demonstrated in the synthetic signal test.

\section{REMARKS AND CONCLUSIONS}

Our analysis focuses on an initial comparison between two different instruments: GNSS stations, which measure displacements, and GINGERINO, which measures rotations. We have developed an analysis approach capable of determining rotations for a constellation of GNSS satellites using two different methods. From this comparison, we found that if we do not remove the influence of tides from the GINGERINO signal, their contribution overwhelms other structures that only become apparent when we also consider tides in the RLG signal, similar to how we handled the stations data.\\
Concerning the coherence structure with a period of 20-60 days, it may be attributed to a phenomenon acting vertically on the ground. This observation is evident only in the Curl method, which, unlike the simple rotation vector around vertical axis, incorporates the vertical component. One potential factor might involve the alteration of pressure phenomena within atmospheric loading. The reconstruction of the GNSS stations signal is affected by these pressure variations, and GINGERINO, being influenced by pressure differences exerted on the ground, experiences deformations in its surrounding area. Additionally, these cycles are also impacted by the presence of the aquifer in the Gran Sasso \cite{DeLuca2018}, which will be further investigated in future studies.\\
The analysis enables us to detect signals with amplitudes one order of magnitude lower than the average amplitudes of the analyzed signal ($10^{-12} rad/s$). In the GINGERINO RLG apparatus, being a single RLG with vertical area vector, it is not possible to separate rotations and tilts of the structure, see variable $\theta$ in eq. \ref{eq:uno}, which depend on the latitude, but also on the orientation of the cavity with respect to the rotation axis. Moreover, being based on a simple mechanical structure, it is limited by rotations of instrumental origin, due to the fact the the laser cavity frame is not rigid enough. Based on this experience a new design has been developed to make tighter the cavity and avoid rotations of instrumental origin \cite{AA31}\\  
At present a more powerful apparatus, called GINGER, is under construction. It involves the implementation of two larger Ring Laser Gyroscopes (RLGs). One of these RLGs will operate at maximum Sagnac signal, called RLX, so with area vector parallel to the Earth rotation axis. For its special alignment, RLX will allow the identification of the orientation of the others RLG of the array, allowing the separation of tilts and rotations. The expectation is to reach 1 part in $10^{11}$ of the Earth rotation rate.\\
In this article, we have focused on a one-year data, as it represents the longest period of available data from GINGERINO. \\ 
In the future, similarly to GNSS stations, multi-years data will be available and we hope  to further investigate and understand the phenomenon. By incorporating GNSS stations, we can gain a better understanding of the surrounding area of the RLG and improve our ability to identify signals that are not solely related to rotation. Ultimately, this will lead to a substantial improvement in the signal-to-noise ratio, enhancing the overall accuracy of our findings. \\
The positive results of the coherence effectively validate the analysis procedure of the RLG which is necessary to eliminate the Laser disturbance, which poses severe limitation for period above 1000s.

\section*{Appendix A: Error Propagation} 


Following are all the equations to derive angular velocity in the rotation vector method. Starting from the linear velocities of the individual stations, the angles defined previously to derive the perpendicular component of the distance vector and the absolute value of the distance itself. 
\begin{equation}
    \omega_1 = \frac{|v_1|}{|r_1|}\sin{(\alpha_1-\theta_1)} 
    \label{w1a}
\end{equation}
In order to do the weighted sum of the individual stations, we must propagate the error in the appropriate way:
\begin{equation}
\begin{aligned}
    \sigma_{\omega_1}^2 &= \left(\frac{\partial\omega_1}{\partial v_1}\right)^2\sigma_{v_1}^2+\left(\frac{\partial\omega_1}{\partial r_1}\right)^2\sigma_{r_1}^2 +\\
    &+\left(\frac{\partial\omega_1}{\partial \alpha_1}\right)^2\sigma_{\alpha_1}^2+\left(\frac{\partial\omega_1}{\partial\theta_1}\right)^2\sigma_{\theta_1}^2 
    \label{sw1}
\end{aligned}
\end{equation}

\begin{align}
    \left(\frac{\partial\omega_1}{\partial v_1}\right)^2 &= \left(\frac{\sin{(\alpha_1-\theta_1)}}{|r_1|}\right)^2 \\
    \left(\frac{\partial\omega_1}{\partial r_1}\right)^2 &= \left(-\frac{|v_1|\sin{(\alpha_1-\theta_1)}}{|r_1|^2}\right)^2 \\
    \left(\frac{\partial\omega_1}{\partial \alpha_1}\right)^2 &= \left(\frac{|v_1|\cos{(\alpha_1-\theta_1)}}{|r_1|}\right)^2 \\
    \left(\frac{\partial\omega_1}{\partial \theta_1}\right)^2 &= \left(-\frac{|v_1|\cos{(\alpha_1-\theta_1)}}{|r_1|}\right)^2
    \label{dw}
\end{align}

\begin{equation}
    \sigma_{v_1} = \sqrt{\frac{v_x^2\sigma_{v_x}^2+v_y^2\sigma_{v_y}^2}{v_x^2+v_y^2}}
    \label{sv1}
\end{equation}

\begin{equation}
    \sigma_{\alpha_1} = \frac{\sqrt{v_y^2\sigma_{v_x}^2+v_x^2\sigma_{v_y}^2}}{v_x^2+v_y^2}
    \label{saplha1}
\end{equation}

\begin{equation}
    \sigma_{v_x} = \sqrt{\sigma_{x_i}^2+\sigma_{x_{i+1}}^2}
    \label{svx}
\end{equation}

all other parts can be obtained analytically from GNSS measurements, instead, $\sigma_{r_1}$ and $\sigma_{\theta_1}$ are evaluated with a parametric bootstrap method \cite{boos2003introduction}, because they are obtained with the “distance” function of Matlab. \\
In our approach, we began by generating normal functions that were centered around the positions of the stations. The width of these functions was determined by the sigma value obtained from our measurements. Using the 'distance' function, we then calculated the distance to GINGERINO as well as the corresponding angle. We repeated this procedure a million times, randomly extracting values from the aforementioned normal distribution at each iteration. This allowed us to obtain distributions of the measurements, and we associated the standard deviations of these distributions with the errors in the distance $r_1$ and angle $ \theta_1 $ . 

\section*{Acknowledgements}
The authors express their profound appreciation to Prof. Giancarlo Cella and Prof. Antonello Ortolan for their stimulating and enlightening discussions, which significantly improved the quality of this article. Their valuable insights and thoughtful feedback played an instrumental role in shaping our overall understanding of the subject matter.
Additionally, special thanks are due to PhD students Alessandro Porcelli and Michele Cardelli, whose thought-provoking questions and curiosities greatly contributed to the clarity of fundamental steps in this analysis. Their input has been invaluable to the development of this work.
We also thank Aladino Govoni and Gaetano De Luca for their contribution to making GINGERINO operational.


\bibliography{bibGINGER_GDS.bib}

\begin{thebibliography}{25}
\expandafter\ifx\csname natexlab\endcsname\relax\def\natexlab#1{#1}\fi
\expandafter\ifx\csname bibnamefont\endcsname\relax
  \def\bibnamefont#1{#1}\fi
\expandafter\ifx\csname bibfnamefont\endcsname\relax
  \def\bibfnamefont#1{#1}\fi
\expandafter\ifx\csname citenamefont\endcsname\relax
  \def\citenamefont#1{#1}\fi
\expandafter\ifx\csname url\endcsname\relax
  \def\url#1{\texttt{#1}}\fi
\expandafter\ifx\csname urlprefix\endcsname\relax\def\urlprefix{URL }\fi
\providecommand{\bibinfo}[2]{#2}
\providecommand{\eprint}[2][]{\url{#2}}

\bibitem[{\citenamefont{{Angela D.V. Di Virgilio, et al.}}(2023)}]{GINGER}
\bibinfo{author}{\bibnamefont{{Angela D.V. Di Virgilio, et al.}}},
  \bibinfo{journal}{to be published in MEMOCS}  (\bibinfo{year}{2023}),
  \urlprefix\url{arXiv:2209.09328}.

\bibitem[{\citenamefont{Belfi et~al.}(2017)}]{GA5}
\bibinfo{author}{\bibfnamefont{J.}~\bibnamefont{Belfi}} \bibnamefont{et~al.},
  \bibinfo{journal}{Rev. Sci. Instrum.} \textbf{\bibinfo{volume}{88}},
  \bibinfo{pages}{034502} (\bibinfo{year}{2017}), \eprint{1702.02789}.

\bibitem[{\citenamefont{Kreemer et~al.}(2014)\citenamefont{Kreemer, Blewitt,
  and Klein}}]{KREEMER14}
\bibinfo{author}{\bibfnamefont{C.}~\bibnamefont{Kreemer}},
  \bibinfo{author}{\bibfnamefont{G.}~\bibnamefont{Blewitt}}, \bibnamefont{and}
  \bibinfo{author}{\bibfnamefont{E.}~\bibnamefont{Klein}},
  \bibinfo{journal}{Geochem. Geophys. Geosyst.} \textbf{\bibinfo{volume}{15}},
  \bibinfo{pages}{3849–3889} (\bibinfo{year}{2014}),
  \urlprefix\url{https://doi.org/10.1002/2014GC005407}.

\bibitem[{\citenamefont{Altamimi et~al.}(2023)\citenamefont{Altamimi,
  Rebischung, Collilieux, Métivier, and Chanard}}]{ALTAMIMI23}
\bibinfo{author}{\bibfnamefont{Z.}~\bibnamefont{Altamimi}},
  \bibinfo{author}{\bibfnamefont{P.}~\bibnamefont{Rebischung}},
  \bibinfo{author}{\bibfnamefont{X.}~\bibnamefont{Collilieux}},
  \bibinfo{author}{\bibfnamefont{L.}~\bibnamefont{Métivier}},
  \bibnamefont{and} \bibinfo{author}{\bibfnamefont{K.}~\bibnamefont{Chanard}},
  \bibinfo{journal}{J. Geod.} \textbf{\bibinfo{volume}{97}},
  \bibinfo{pages}{47} (\bibinfo{year}{2023}),
  \urlprefix\url{https://doi.org/10.1007/s00190-023-01738-w}.

\bibitem[{\citenamefont{Okazaki et~al.}(2021)\citenamefont{Okazaki, Fukahata,
  and Nishimura}}]{OKAZAKI21}
\bibinfo{author}{\bibfnamefont{T.}~\bibnamefont{Okazaki}},
  \bibinfo{author}{\bibfnamefont{Y.}~\bibnamefont{Fukahata}}, \bibnamefont{and}
  \bibinfo{author}{\bibfnamefont{T.}~\bibnamefont{Nishimura}},
  \bibinfo{journal}{Earth Planets Space} \textbf{\bibinfo{volume}{73}},
  \bibinfo{pages}{153} (\bibinfo{year}{2021}),
  \urlprefix\url{https://doi.org/10.1186/s40623-021-01474-5}.

\bibitem[{\citenamefont{Serpelloni et~al.}(2022)\citenamefont{Serpelloni,
  Cavaliere, Martelli, Pintori, Anderlini, Borghi, Randazzo, Bruni, Devoti,
  Perfetti et~al.}}]{SERPELLONI22}
\bibinfo{author}{\bibfnamefont{E.}~\bibnamefont{Serpelloni}},
  \bibinfo{author}{\bibfnamefont{A.}~\bibnamefont{Cavaliere}},
  \bibinfo{author}{\bibfnamefont{L.}~\bibnamefont{Martelli}},
  \bibinfo{author}{\bibfnamefont{F.}~\bibnamefont{Pintori}},
  \bibinfo{author}{\bibfnamefont{L.}~\bibnamefont{Anderlini}},
  \bibinfo{author}{\bibfnamefont{A.}~\bibnamefont{Borghi}},
  \bibinfo{author}{\bibfnamefont{D.}~\bibnamefont{Randazzo}},
  \bibinfo{author}{\bibfnamefont{S.}~\bibnamefont{Bruni}},
  \bibinfo{author}{\bibfnamefont{R.}~\bibnamefont{Devoti}},
  \bibinfo{author}{\bibfnamefont{P.}~\bibnamefont{Perfetti}},
  \bibnamefont{et~al.}, \bibinfo{journal}{Front. Earth Sci.}
  \textbf{\bibinfo{volume}{10}}, \bibinfo{pages}{907897}
  (\bibinfo{year}{2022}),
  \urlprefix\url{https://doi.org/10.3389/feart.2022.907897}.

\bibitem[{\citenamefont{Palano et~al.}(2023)\citenamefont{Palano, Calcaterra,
  Gambino, Porfidia, and Sparacino}}]{PALANO23}
\bibinfo{author}{\bibfnamefont{m.}~\bibnamefont{Palano}},
  \bibinfo{author}{\bibfnamefont{S.}~\bibnamefont{Calcaterra}},
  \bibinfo{author}{\bibfnamefont{P.}~\bibnamefont{Gambino}},
  \bibinfo{author}{\bibfnamefont{B.}~\bibnamefont{Porfidia}}, \bibnamefont{and}
  \bibinfo{author}{\bibfnamefont{F.}~\bibnamefont{Sparacino}},
  \bibinfo{journal}{Results Geophys. Sci.} \textbf{\bibinfo{volume}{14}},
  \bibinfo{pages}{100056} (\bibinfo{year}{2023}),
  \urlprefix\url{https://doi.org/10.1016/j.ringps.2023.100056}.

\bibitem[{\citenamefont{Leone et~al.}(2023)\citenamefont{Leone, D’Agostino,
  Esposito, and Fiorillo}}]{LEONE23}
\bibinfo{author}{\bibfnamefont{G.}~\bibnamefont{Leone}},
  \bibinfo{author}{\bibfnamefont{N.}~\bibnamefont{D’Agostino}},
  \bibinfo{author}{\bibfnamefont{L.}~\bibnamefont{Esposito}}, \bibnamefont{and}
  \bibinfo{author}{\bibfnamefont{F.}~\bibnamefont{Fiorillo}},
  \bibinfo{journal}{Environmental Earth Sciences}
  \textbf{\bibinfo{volume}{82}}, \bibinfo{pages}{240} (\bibinfo{year}{2023}),
  \urlprefix\url{https://doi.org/10.1007/s12665-023-10905-3}.

\bibitem[{\citenamefont{Pintori et~al.}(2021)\citenamefont{Pintori, Serpelloni,
  Longuevergne, Garcia, Faenza, D'Alberto, Gualandi, and
  Belardinelli}}]{PINTORI21}
\bibinfo{author}{\bibfnamefont{F.}~\bibnamefont{Pintori}},
  \bibinfo{author}{\bibfnamefont{E.}~\bibnamefont{Serpelloni}},
  \bibinfo{author}{\bibfnamefont{L.}~\bibnamefont{Longuevergne}},
  \bibinfo{author}{\bibfnamefont{A.}~\bibnamefont{Garcia}},
  \bibinfo{author}{\bibfnamefont{L.}~\bibnamefont{Faenza}},
  \bibinfo{author}{\bibfnamefont{L.}~\bibnamefont{D'Alberto}},
  \bibinfo{author}{\bibfnamefont{A.}~\bibnamefont{Gualandi}}, \bibnamefont{and}
  \bibinfo{author}{\bibfnamefont{M.~E.} \bibnamefont{Belardinelli}},
  \bibinfo{journal}{J. Geoph. Res.} \textbf{\bibinfo{volume}{126}},
  \bibinfo{pages}{e2020JB020586} (\bibinfo{year}{2021}),
  \urlprefix\url{https://doi.org/10.1029/2020JB020586}.

\bibitem[{\citenamefont{Michel et~al.}(2021)\citenamefont{Michel,
  Santamaría-Gómez, Boy, Perosanz, and Loyer}}]{MICHEL21}
\bibinfo{author}{\bibfnamefont{A.}~\bibnamefont{Michel}},
  \bibinfo{author}{\bibfnamefont{A.}~\bibnamefont{Santamaría-Gómez}},
  \bibinfo{author}{\bibfnamefont{J.-P.} \bibnamefont{Boy}},
  \bibinfo{author}{\bibfnamefont{F.}~\bibnamefont{Perosanz}}, \bibnamefont{and}
  \bibinfo{author}{\bibfnamefont{S.}~\bibnamefont{Loyer}},
  \bibinfo{journal}{Remote Sens.} \textbf{\bibinfo{volume}{13}},
  \bibinfo{pages}{4523} (\bibinfo{year}{2021}),
  \urlprefix\url{https://doi.org/10.1007/s10291-020-0959-3}.

\bibitem[{\citenamefont{Devoti et~al.}(2015)\citenamefont{Devoti, Zuliani,
  Braitenberg, Fabris, and Grillo}}]{DEVOTI15}
\bibinfo{author}{\bibfnamefont{R.}~\bibnamefont{Devoti}},
  \bibinfo{author}{\bibfnamefont{D.}~\bibnamefont{Zuliani}},
  \bibinfo{author}{\bibfnamefont{C.}~\bibnamefont{Braitenberg}},
  \bibinfo{author}{\bibfnamefont{P.}~\bibnamefont{Fabris}}, \bibnamefont{and}
  \bibinfo{author}{\bibfnamefont{B.}~\bibnamefont{Grillo}},
  \bibinfo{journal}{Earth and Planetary Science Letters}
  \textbf{\bibinfo{volume}{419}}, \bibinfo{pages}{134} (\bibinfo{year}{2015}),
  \urlprefix\url{https://doi.org/10.1016/j.epsl.2015.03.023}.

\bibitem[{\citenamefont{Mémin et~al.}(2020)\citenamefont{Mémin, Boy, and
  Santamaria-Gomez}}]{MEMIN20}
\bibinfo{author}{\bibfnamefont{A.}~\bibnamefont{Mémin}},
  \bibinfo{author}{\bibfnamefont{J.}~\bibnamefont{Boy}}, \bibnamefont{and}
  \bibinfo{author}{\bibfnamefont{A.}~\bibnamefont{Santamaria-Gomez}},
  \bibinfo{journal}{GPS Solut.} \textbf{\bibinfo{volume}{24}},
  \bibinfo{pages}{45} (\bibinfo{year}{2020}),
  \urlprefix\url{https://doi.org/10.1007/s10291-020-0959-3}.

\bibitem[{\citenamefont{Martens et~al.}(2020)\citenamefont{Martens, Argus,
  Norberg, Blewitt, Herring, Moore, Hammond, and Kreemer}}]{MARTENS20}
\bibinfo{author}{\bibfnamefont{H.}~\bibnamefont{Martens}},
  \bibinfo{author}{\bibfnamefont{D.}~\bibnamefont{Argus}},
  \bibinfo{author}{\bibfnamefont{C.}~\bibnamefont{Norberg}},
  \bibinfo{author}{\bibfnamefont{G.}~\bibnamefont{Blewitt}},
  \bibinfo{author}{\bibfnamefont{T.}~\bibnamefont{Herring}},
  \bibinfo{author}{\bibfnamefont{A.}~\bibnamefont{Moore}},
  \bibinfo{author}{\bibfnamefont{W.}~\bibnamefont{Hammond}}, \bibnamefont{and}
  \bibinfo{author}{\bibfnamefont{C.}~\bibnamefont{Kreemer}},
  \bibinfo{journal}{J. Geod.} \textbf{\bibinfo{volume}{94}},
  \bibinfo{pages}{115} (\bibinfo{year}{2020}),
  \urlprefix\url{https://doi.org/10.1007/s00190-020-01445-w}.

\bibitem[{\citenamefont{Tercjak and Brzezi{\'n}ski}(2017)}]{diciotto}
\bibinfo{author}{\bibfnamefont{M.}~\bibnamefont{Tercjak}} \bibnamefont{and}
  \bibinfo{author}{\bibfnamefont{A.}~\bibnamefont{Brzezi{\'n}ski}},
  \bibinfo{journal}{Pure and Applied Geophysics}
  \textbf{\bibinfo{volume}{174}}, \bibinfo{pages}{2719} (\bibinfo{year}{2017}).

\bibitem[{\citenamefont{Di~Virgilio et~al.}(2017)\citenamefont{Di~Virgilio,
  Belfi, Ni, Beverini, Carelli, Maccioni, and Porzio}}]{dieci}
\bibinfo{author}{\bibfnamefont{A.~D.~V.} \bibnamefont{Di~Virgilio}},
  \bibinfo{author}{\bibfnamefont{J.}~\bibnamefont{Belfi}},
  \bibinfo{author}{\bibfnamefont{W.-T.} \bibnamefont{Ni}},
  \bibinfo{author}{\bibfnamefont{N.}~\bibnamefont{Beverini}},
  \bibinfo{author}{\bibfnamefont{G.}~\bibnamefont{Carelli}},
  \bibinfo{author}{\bibfnamefont{E.}~\bibnamefont{Maccioni}}, \bibnamefont{and}
  \bibinfo{author}{\bibfnamefont{A.}~\bibnamefont{Porzio}},
  \bibinfo{journal}{Eur. Phys. J. Plus} \textbf{\bibinfo{volume}{132}},
  \bibinfo{pages}{157} (\bibinfo{year}{2017}).

\bibitem[{\citenamefont{Di~Virgilio et~al.}(2019)\citenamefont{Di~Virgilio,
  Beverini, Carelli, Ciampini, Fuso, and Maccioni}}]{AA26}
\bibinfo{author}{\bibfnamefont{A.~D.~V.} \bibnamefont{Di~Virgilio}},
  \bibinfo{author}{\bibfnamefont{N.}~\bibnamefont{Beverini}},
  \bibinfo{author}{\bibfnamefont{G.}~\bibnamefont{Carelli}},
  \bibinfo{author}{\bibfnamefont{D.}~\bibnamefont{Ciampini}},
  \bibinfo{author}{\bibfnamefont{F.}~\bibnamefont{Fuso}}, \bibnamefont{and}
  \bibinfo{author}{\bibfnamefont{E.}~\bibnamefont{Maccioni}},
  \bibinfo{journal}{Eur. Phys. J. C} \textbf{\bibinfo{volume}{79}},
  \bibinfo{pages}{573} (\bibinfo{year}{2019}), \eprint{1904.02533}.

\bibitem[{\citenamefont{Di~Virgilio et~al.}(2020)\citenamefont{Di~Virgilio,
  Basti, Beverini, Bosi, Carelli, Ciampini, Fuso, Giacomelli, Maccioni, Marsili
  et~al.}}]{PRR2020}
\bibinfo{author}{\bibfnamefont{A.~D.~V.} \bibnamefont{Di~Virgilio}},
  \bibinfo{author}{\bibfnamefont{A.}~\bibnamefont{Basti}},
  \bibinfo{author}{\bibfnamefont{N.}~\bibnamefont{Beverini}},
  \bibinfo{author}{\bibfnamefont{F.}~\bibnamefont{Bosi}},
  \bibinfo{author}{\bibfnamefont{G.}~\bibnamefont{Carelli}},
  \bibinfo{author}{\bibfnamefont{D.}~\bibnamefont{Ciampini}},
  \bibinfo{author}{\bibfnamefont{F.}~\bibnamefont{Fuso}},
  \bibinfo{author}{\bibfnamefont{U.}~\bibnamefont{Giacomelli}},
  \bibinfo{author}{\bibfnamefont{E.}~\bibnamefont{Maccioni}},
  \bibinfo{author}{\bibfnamefont{P.}~\bibnamefont{Marsili}},
  \bibnamefont{et~al.}, \bibinfo{journal}{Phys. Rev. Res.}
  \textbf{\bibinfo{volume}{2}}, \bibinfo{pages}{032069} (\bibinfo{year}{2020}),
  \urlprefix\url{https://link.aps.org/doi/10.1103/PhysRevResearch.2.032069}.

\bibitem[{\citenamefont{{Di Virgilio, Angela D.} et~al.}(2021)\citenamefont{{Di
  Virgilio, Angela D.}, {Altucci, Carlo}, {Bajardi, Francesco}, {Basti,
  Andrea}, {Beverini, Nicol\`o}, {Capozziello, Salvatore}, {Carelli, Giorgio},
  {Ciampini, Donatella}, {Fuso, Francesco}, {Giacomelli, Umberto}
  et~al.}}]{EPJC2020}
\bibinfo{author}{\bibnamefont{{Di Virgilio, Angela D.}}},
  \bibinfo{author}{\bibnamefont{{Altucci, Carlo}}},
  \bibinfo{author}{\bibnamefont{{Bajardi, Francesco}}},
  \bibinfo{author}{\bibnamefont{{Basti, Andrea}}},
  \bibinfo{author}{\bibnamefont{{Beverini, Nicol\`o}}},
  \bibinfo{author}{\bibnamefont{{Capozziello, Salvatore}}},
  \bibinfo{author}{\bibnamefont{{Carelli, Giorgio}}},
  \bibinfo{author}{\bibnamefont{{Ciampini, Donatella}}},
  \bibinfo{author}{\bibnamefont{{Fuso, Francesco}}},
  \bibinfo{author}{\bibnamefont{{Giacomelli, Umberto}}}, \bibnamefont{et~al.},
  \bibinfo{journal}{Eur. Phys. J. C} \textbf{\bibinfo{volume}{81}},
  \bibinfo{pages}{400} (\bibinfo{year}{2021}),
  \urlprefix\url{https://doi.org/10.1140/epjc/s10052-021-09199-1}.

\bibitem[{\citenamefont{{Di Virgilio, A. D. V.} et~al.}(2022)\citenamefont{{Di
  Virgilio, A. D. V.}, {Terreni, G.}, {Basti, A.}, {Beverini, N.}, {Carelli,
  G.}, {Ciampini, D.}, {Fuso, F.}, {Maccioni, E.}, {Marsili, P.}, {Kodet, J.}
  et~al.}}]{EPJC2021}
\bibinfo{author}{\bibnamefont{{Di Virgilio, A. D. V.}}},
  \bibinfo{author}{\bibnamefont{{Terreni, G.}}},
  \bibinfo{author}{\bibnamefont{{Basti, A.}}},
  \bibinfo{author}{\bibnamefont{{Beverini, N.}}},
  \bibinfo{author}{\bibnamefont{{Carelli, G.}}},
  \bibinfo{author}{\bibnamefont{{Ciampini, D.}}},
  \bibinfo{author}{\bibnamefont{{Fuso, F.}}},
  \bibinfo{author}{\bibnamefont{{Maccioni, E.}}},
  \bibinfo{author}{\bibnamefont{{Marsili, P.}}},
  \bibinfo{author}{\bibnamefont{{Kodet, J.}}}, \bibnamefont{et~al.},
  \bibinfo{journal}{Eur. Phys. J. C} \textbf{\bibinfo{volume}{82}},
  \bibinfo{pages}{824} (\bibinfo{year}{2022}),
  \urlprefix\url{https://doi.org/10.1140/epjc/s10052-022-10798-9}.

\bibitem[{IER()}]{IERS}
\emph{\bibinfo{title}{Iers data: Compute time series of the earth orientation
  matrix / quaternion or derived parameters between two dates}},
  \urlprefix\url{https://hpiers.obspm.fr/eop-pc/index.php?index=rotation&lang=en}.

\bibitem[{GOT()}]{GOTIC2}
\emph{\bibinfo{title}{Gotic2$\_{Mod}$: Modified gotic2, software for estimation
  of ocean tidal loading effects at subsurface observation}},
  \urlprefix\url{https://www.gsj.jp/data/openfile/no0705/openfile0705indexEN.html}.

\bibitem[{\citenamefont{Allmendinger et~al.}(2007)\citenamefont{Allmendinger,
  Reilinger, and Loveless}}]{Allmendinger}
\bibinfo{author}{\bibfnamefont{R.}~\bibnamefont{Allmendinger}},
  \bibinfo{author}{\bibfnamefont{R.}~\bibnamefont{Reilinger}},
  \bibnamefont{and} \bibinfo{author}{\bibfnamefont{J.}~\bibnamefont{Loveless}},
  \bibinfo{journal}{Tectonics} \textbf{\bibinfo{volume}{26}}
  (\bibinfo{year}{2007}).

\bibitem[{\citenamefont{De~Luca et~al.}(2018)\citenamefont{De~Luca, Di~Carlo,
  and Tallini}}]{DeLuca2018}
\bibinfo{author}{\bibfnamefont{G.}~\bibnamefont{De~Luca}},
  \bibinfo{author}{\bibfnamefont{G.}~\bibnamefont{Di~Carlo}}, \bibnamefont{and}
  \bibinfo{author}{\bibfnamefont{M.}~\bibnamefont{Tallini}},
  \bibinfo{journal}{Scientific Reports} \textbf{\bibinfo{volume}{8}},
  \bibinfo{pages}{15982} (\bibinfo{year}{2018}), ISSN
  \bibinfo{issn}{2045-2322},
  \urlprefix\url{https://doi.org/10.1038/s41598-018-34444-1}.

\bibitem[{\citenamefont{Basti et~al.}(2021)}]{AA31}
\bibinfo{author}{\bibfnamefont{A.}~\bibnamefont{Basti}} \bibnamefont{et~al.},
  \bibinfo{journal}{Eur. Phys. J. Plus} \textbf{\bibinfo{volume}{136}},
  \bibinfo{pages}{537} (\bibinfo{year}{2021}), \eprint{2101.07686}.

\bibitem[{\citenamefont{Boos}(2003)}]{boos2003introduction}
\bibinfo{author}{\bibfnamefont{D.~D.} \bibnamefont{Boos}},
  \bibinfo{journal}{Statistical science} \textbf{\bibinfo{volume}{18}},
  \bibinfo{pages}{168} (\bibinfo{year}{2003}).

\end{thebibliography}

\end{document}